\documentclass[12pt,amsmath,amssymb]{iopart}

\usepackage{graphicx}
\usepackage{color}
\usepackage{array}
\usepackage{braket}
\usepackage{caption}
\usepackage[cp1251]{inputenc}
\usepackage[english]{babel}

\begin{document}

\title{Schmidt decomposition for non-collinear biphoton angular wave functions}
\author{M.V. Fedorov}
\address{A M Prokhorov General Physics Institute, Russian Academy of Science, Moscow, Russia.\\
Moscow Institute of Physics and Technology, Dolgoprudny, Moscow Region, Russia}
\begin{abstract}
Schmidt modes of non-collinear biphoton angular wave functions are found analytically. The experimentally realizable procedure is described for their separation. Parameters of the Schmidt decomposition are used for evaluation of the degree of biphoton's angular entanglement.
\end{abstract}

\maketitle

\section{Introduction}
As known, the Schmidt decomposition is a powerful instrument for analysis of correlations (entanglement) of pure bipartite states \cite{CP}. The most often met systems of such kind are states of two photons produced in processes of Spontaneous Parametric Down-Conversion (SPDC), in which some photons of the pump decay in a nonlinear crystal for pairs of photons of smaller frequencies. Regimes of SPDC depend on features of the pump and used nonlinear crystals. The simplest collinear degenerate regime is that of a plane monochromatic wave giving rise to collinearly propagating SPDC photons with coinciding frequencies equal to the half of the pump frequency. In this case the only degree of freedom of SPDC photons in which they can be entangled or not is their polarization. Such states are known as polarization biphoton qutrits. Their features have been investigated by many authors in a number of papers, for example, such as \cite{Burl-Kly,Bechmann,Pask-You,Burl-Che,Lanyon,NJP,Che}. In particular, the Schmidt decomposition of biphoton polarization qutrits was discussed in the works \cite{CP,Che,RocPro} and, as a reminder, some of these results are reproduced briefly in the following section. In the same collinear and degenerate regime but with transversely spreading pump and SPDC photon beams the angular Schmidt modes and decomposition were found experimentally in  the works \cite{Ku-Stra-1,Ku-Stra-2}. In section 3 we will describe the Schmidt decomposition and the method of separation of the angular Schmidt modes for the non-collinear degenerate regime of SPDC but with ignored spreading of the pump and SPDC photons. At last, section 4 is devoted to the derivation of the angular Schmidt modes in the same non-collinear degenerate regime but with spearing of photon beams completely taken into account. The derivation will be strongly related to and based on the proposal of a scheme in which the Schmidt modes can be found and separated experimentally.

\section{Polarization biphoton qutrits}
The most general form of the state vector of biphoton polarization qutrits is given by
left
\begin{equation}
 \label{qutrit}
 \ket{\Psi}=C_1\ket{2_H}+C_2\ket{1_H,1_V}+C_3\ket{2_V},
\end{equation}
where $H$ and $V$ refer to the horizontal and vertical polarizations of photons, and $C_{1,2,3}$ are arbitrary complex constants obeying the normalization condition $|C_1|^2+|C_2|^2+|C_3|^2=1$. As shown \cite{CP,Che}, at any values of the constants $C_i$ state vectors of biphoton polarization qutrits can be presented in the Schmidt-decomposition form
\begin{equation}
 \label{Schm-dec-qutrit}
 \ket{\Psi}=\sqrt{\lambda_+}\ket{2_+}+\sqrt{\lambda_-}\ket{2_-},
\end{equation}
where $\ket{2_+}$ and $\ket{2_-}$ are state vectors of two photons in two orthogonal Schmidt modes $\ket{1_+}$ and $\ket{1_-}$. There are several ways of finding explicitly Schmidt modes in terms of the constants $C_i$ \cite{CP,Che,RocPro}. The conventional method \cite{NJP} is related to the use of polarization wave functions  $\Psi(\sigma_1,\sigma_2)=\braket{\sigma_1,\sigma_2|\Psi}$ where $\sigma_1$ and $\sigma_2$ are polarization variables of two photons such that $\braket{\sigma_{1,2}|1_H}=\delta_{\sigma_{1,2},H}$ and $\braket{\sigma_{1,2}|1_V}=\delta_{\sigma_{1,2},V}$, construction of the total density matrix $\Psi(\sigma_1,\sigma_2)\Psi^*(\sigma_1^\prime,\sigma_2^\prime)$, and its reduction over one of the variables, $\sigma_2=\sigma_2^\prime$ or $\sigma_1=\sigma_1^\prime$. Eigenfunctions and eigenvalues of the reduced density matrices are just the Schmidt modes and parameters $\lambda_+$ and $\lambda_-$ determining the Schmidt decomposition (\ref{Schm-dec-qutrit}). These parameters obey the normalization condition $\lambda_++\lambda_-=1$, and their variation intervals are $1\geq\lambda_+\geq 0.5$ and $0.5\geq\lambda_-\geq 0$. As was shown \cite{NJP}, these parameter determine completely the qutrit's degree of polarization $P$ and such entanglement quantifiers as the concurrence $C$, Schmidt number $K$ and entropy of the reduced states
\begin{equation}
 \label{degrees}
 \begin{array}{l}
 P=\lambda_+-\lambda_-,\;K=\displaystyle\frac{1}{\lambda_+^2+\lambda_-^2},\\
 C=2\sqrt{\lambda_+\lambda_-}\\
 S_r=-\log_2\lambda_+-\log_2\lambda_-.
 \end{array}
\end{equation}
Qutrits are maximally entangled in the case $\lambda_+=\lambda_-=0.5$ and disentangled when $\lambda_+=1,\,\lambda_-=0$.

\begin{figure}[h]
\includegraphics[width=6cm]{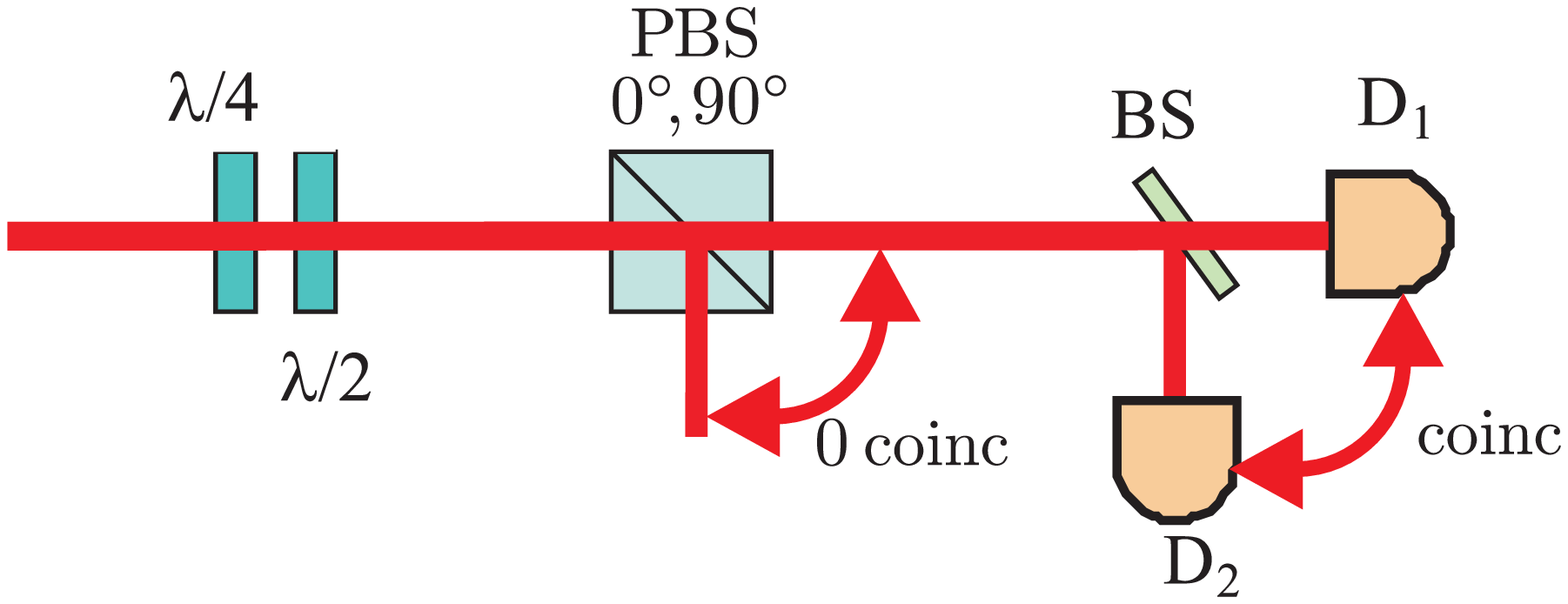}
\caption{{\protect\footnotesize {Scheme of experiments for separation of Schmidt modes of biphoton polarization qutrits, $\lambda/4$ and $\lambda/2$ are quarter- and half-wavelength plates, ${\rm PBS}$ with labels $0^\circ$ and $90^\circ$ denote Polarization Beam Splitter installed, correspondingly, conventionally or turned for $90^\circ$ around the propagation axis, ${\rm BS}$ denotes a non-polarizing Beam Splitter, $D_1$ and $D_2$ are detectors.}}}
\label{Fig1}
\end{figure}

In the paper \cite{Che} we have outlined the method of direct experimental measurement of the parameters $\lambda_+$ and $\lambda_-$. The scheme of such experiment is shown in Fig. \ref{Fig1}.
The first step of this experiment should consist in the transformation of polarizations of the orthogonal Schmidt modes $\ket{1_+}$ and $\ket{1_-}$, correspondingly, to the horizontal and vertical ones to reduce the Schmidt decomposition (\ref{Schm-dec-qutrit}) to the simplest form
\begin{equation}
 \label{Schm-H-V}
 \ket{\Psi}=\sqrt{\lambda_+}\ket{2_H}+e^{2i\phi}\sqrt{\lambda_-}\ket{2_V},
\end{equation}
where $\phi$ is some phase, which can be easily changed to be used for encoding information \cite{Che} but which does not affect the degree of entanglement of the state (\ref{Schm-dec-qutrit}), (\ref{Schm-H-V}). In experiment the transformation (\ref{Schm-dec-qutrit}) $\rightarrow$ (\ref{Schm-H-V}) can be provided by appropriately installed half- and quarter-wavelength plates. Correct orientation of these plates can be found experimentally  from the condition of zero coincidence signal between two channels immediately after the Polarization Beam Splitter (PBS) as shown in Fig. \ref{Fig1}. Under this condition after PBS one gets two beams containing pairs of separated Schmidt modes: $\ket{2_H}$ transmitted and $\ket{2_V}$ reflected or vice versa if the PBS is turned for $90_\circ$ around the original propagation direction. By measuring the relative amounts of such pairs one finds the probabilities of their appearance equal to $\lambda_+$ and $\lambda_-$. Thus, if $N_{HH}$ and $N_{VV}$ are amounts of clicks of the detectors registering horizontally and vertically polarized photons per some given time, the Schmidt-decomposition parameters are given by
\begin{equation}
 \label{lambda via N}
 \lambda_+=\frac{N_{HH}}{N_{HH}+N_{VV}},\quad\lambda_-=\frac{N_{VV}}{N_{HH}+N_{VV}}.
\end{equation}
The described procedure is general and valid for any qutrits (\ref{qutrit}) with arbitrary unknown parameters $C_{1,2,3}$. But in many special cases this is not needed at all or can be significantly simplified. For example, in the case of the state with two photons of different polarizations its Schmidt decomposition can be found by a simple transformation to the basis turned for $45^\circ$:
\begin{equation}
 \label{Schm-decomp-HV}
 \ket{1_H,1_V}=\frac{1}{\sqrt{2}}\Big(\ket{2_{45^\circ}}-\ket{2_{135^\circ}}\Big).
\end{equation}
The right-hand side of this equation is just the Schmidt decomposition with the Schmidt modes $\ket{1_{45^\circ}}$ and $i\ket{1_{135^\circ}}$. For experimental separation of the pairs of these Schmidt modes one needs only the PBS turned for $45^\circ$ around the propagation axis. Then the transmitted and reflected beams will contain pairs of photons in one or another Schmidt mode, i.e., both polarized either in the direction at $45^\circ$ or $135^\circ$ with respect to the horizontal axis. The amounts of pairs to be measured in each channel will be equal to each other, which indicates that $\lambda_+=\lambda_-=0.5$, and the state $\ket{1_H,1_V}$ is maximally entangled \cite{NJP}.

\section{Angular entanglement}
\begin{figure}[h]
\includegraphics[width=6.5cm]{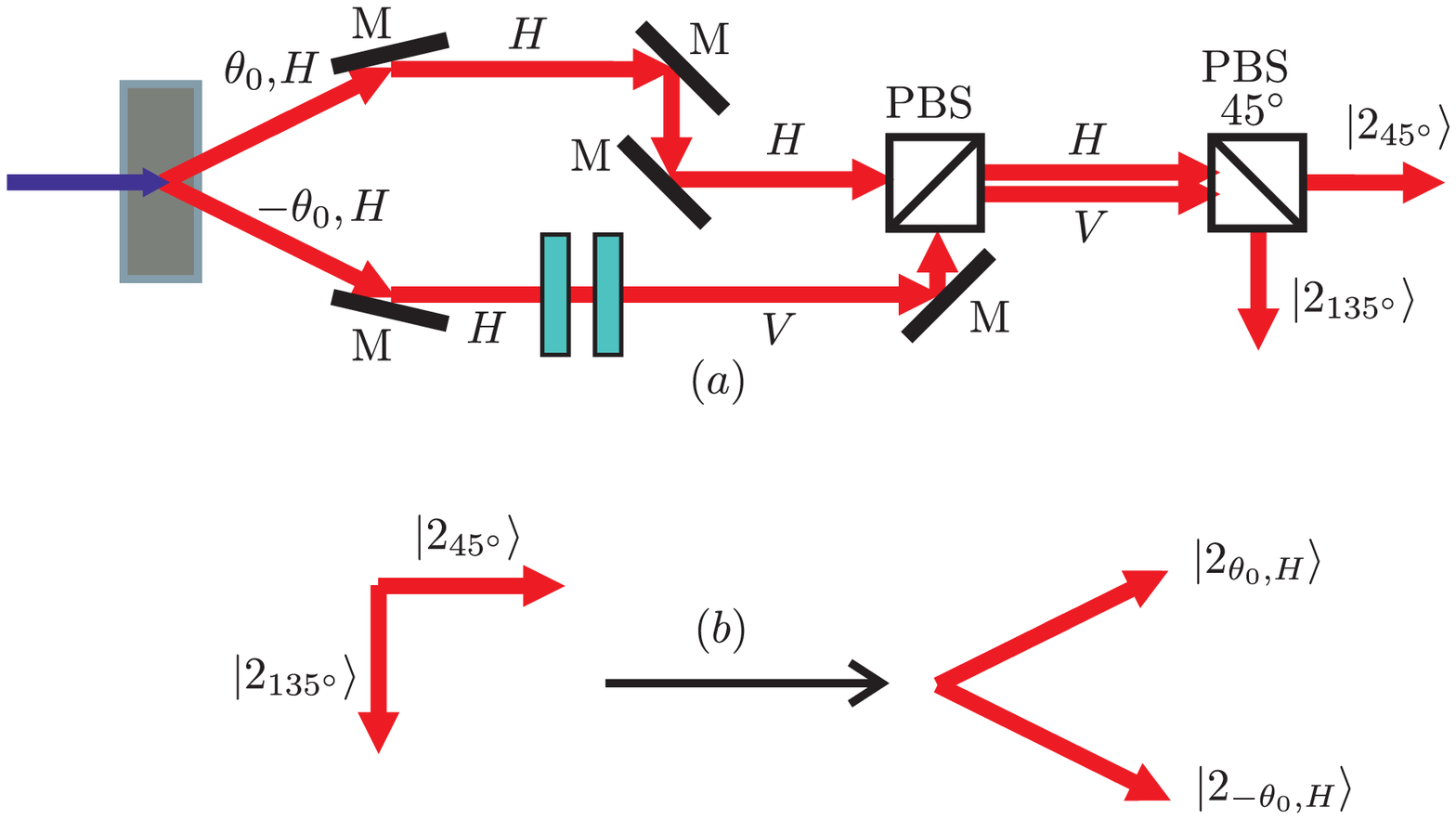}
\caption{{\protect\footnotesize {(a) A scheme for separation of angular Schmidt modes and (b) additional manipulations with separated Schmidt modes; PBS denotes Polarization Beam Splitters, labeling $45^\circ$ refers to the PBS turned for $45^\circ$ around the propagation axis, $M$ are mirrors}}}
\label{Fig2}
\end{figure}

Let us consider now the state of two photons produced in the non-collinear degenerate regime of SPDC with the type-I phase matching. The latter means that polarizations of both photons are horizontal, and there is no polarization entanglement. Let angles of propagation of these photons be  $\theta_0$ and $-\theta_0$. These angles determine two modes in which each of two photons can be found, but they can never appear in one of these modes together. The state vector of such state is $\ket{1_{\theta_0,H},1_{-\theta_0,H}}$. As photons are indistinguishable, localization of each photon in the modes $\theta_0$ and $-\theta_0$ is uncertain and in the same time there is a specific correlation of localization: if one of photons belongs to one of two modes, the second photon with 100$\%$ probability belongs to the other mode. This uncertainty and correlation of localization indicate that the state $\ket{1_{\theta_0,H},1_{-\theta_0,H}}$ is entangled. The problem is in finding a scheme of an experiment in which one could separate pairs of Schmidt modes of this state in a manner similar to that described above for the polarization-entangled state $\ket{1_H,1_V}$ (Eq. \ref{Schm-decomp-HV}). A scheme of such experiment is shown in Fig. \ref{Fig2}.
The first step of this scheme consists of changing the photon polarization in one of two channels $H\rightarrow V$ with the help of $\lambda/4$ and $\lambda/2$ plates (green in Fig. \ref{Fig2}). This change does not affect the degree of entanglement because it does not change the amount of modes accessible for two photon. Still we have two modes, but now they have a double labeling: $\theta_0,H$ and $-\theta_0,V$. The second step is merging two beams into a single one with the help of PBS. After PBS both photons propagate together in the same direction. But, still there are two modes corresponding to different polarizations, $H$ and $V$. As previously, because of indistinguishability of photons, belonging of anyone of them to any given mode remains uncertain and, hence, the system keeps the same degree of entanglement as it had before the performed manipulations. In fact, what is done now is the substitution of the angular entanglement by equivalent polarization entanglement of two photons, for which separation of Schmidt modes is much easier and can be done as described above for the state $\ket{1_H,1_V}$ above. Specifically, the merged state arising after PBS in Fig. \ref{Fig2} has to be sent to the second PBS, turned for $45^\circ$ around the propagation axis and separating photons having polarizations along the directions $45^\circ$ and $135^\circ$ with respect to the horizontal axis. As follows from Eq. (\ref{Schm-decomp-HV}), after the second PBS in Fig. \ref{Fig2} pairs of photons will be either transmitted or reflected depending on their polarization ($45^\circ$ or $135^\circ$) but none of them will be split between two channels. The picture in Fig. \ref{Fig2}$(b)$ shows schematically that, if needed, after separation of pairs of photons one can change both directions of their propagation and polarizations in each channel separately to return to the original geometry of two beams propagating in directions $\theta_0$ and $-\theta_0$ with the same horizontal polarization of all photons. The difference with the original state $\ket{1_{\theta_0,H},1_{-\theta_0,H}}$ is in regrouping photons in such a way that both of them in each SPDC pair propagate now together in one or another direction. Mathematically the arising state is characterized by the state vector in the Schmidt-decomposition form
\begin{equation}
 \label{Schm-angle-nonspeding}
 \ket{\Psi}=\frac{1}{\sqrt{2}}\Big(\ket{2_{\theta_0,H}}-\ket{2_{-\theta_0,H}}\Big).
\end{equation}
By counting amounts of photons in each channel after all manipulations one can find the parameters of the Schmidt decomposition $\lambda_+$ and $\lambda_-$ (\ref{lambda via N}) which have to be close to 0.5 both to indicate that the state of two non-collinear SPDC photons with coinciding polarizations is maximally entangled.

\section{Angular Schmidt modes of non-collinear spreading biphoton beams}

Let us consider now in a more detailed form the structure of photon angular distributions in the same non-collinear frequency-degenerate SPDC regime as in the previous section, but with finite transverse widths of both the pump and SPDC photon beams. In the case of type-I phase matching, under some assumptions, the biphoton angular wave function can be taken in the form
\begin{equation}
 \Psi(\theta_1,\theta_2)=N\exp\left(-\frac{(\theta_1+\theta_2)^2}{2\Delta\theta_p^2}\right)
 \label{wf}
 {\rm sinc}\left(\frac{(\theta_1-\theta_2)^2-4\theta_0^2}{2\Delta\theta_L^2}\right),
\end{equation}
where $N$ is the normalization factor, ${\rm sinc}\,x=\sin x/x$, $\theta_1$ and $\theta_2$ are angles between the wave vectors of two emitted photons and central direction of propagation of the pump ($0z$-axis), $\theta_0$ and $-\theta_0$ are the angles between the central propagation directions of the beams of emitted photons and the $z$-axis, $\Delta\theta_p$ and $\Delta\theta_L$ are, respectively, the angular width of the pump wave and the width of the angular distributions in the the beams of emitted photons related to the finite length of the nonlinear crystal $L$
\begin{equation}
 \Delta\theta_p=\frac{\lambda_p}{\pi n_0d},\,\Delta\theta_L=\sqrt{\frac{2\lambda_p.}{\pi n_0L}},\,
 \label{widths}
 \theta_0=\sqrt{2\frac{n_o-n_e}{n_0}}
\end{equation}
with $n_o$ and $n_e$ being the refracting indices of the ordinary and extraordinary waves in in a crystal in the direction of the $z$-axis and $d$ the waist of the pump wave. The widths and all characteristic angles are assumed to be small:
\begin{equation}
 \label{narrow peaks}
 \Delta\theta_p,\, \Delta\theta_L \ll\theta_0\ll 1.
\end{equation}
Note also that the argument of the sinc-function in Eq. (\ref{wf}) does not contain terms linear in $\theta_1$ and $\theta_2$. This is correct in a general case for measurements in the plane perpendicular to that containing the optical axis of the crystal. For any other planes of measurements the linear terms can be dropped only if the pump width $\Delta\theta_p$ is sufficiently small. Otherwise the linear terms can affect significantly photon distributions, which has been shown and widely discussed for the case of collinear regime in  Refs. \cite{PRL,PRA}.

\begin{figure}[h]
\includegraphics[width=6cm]{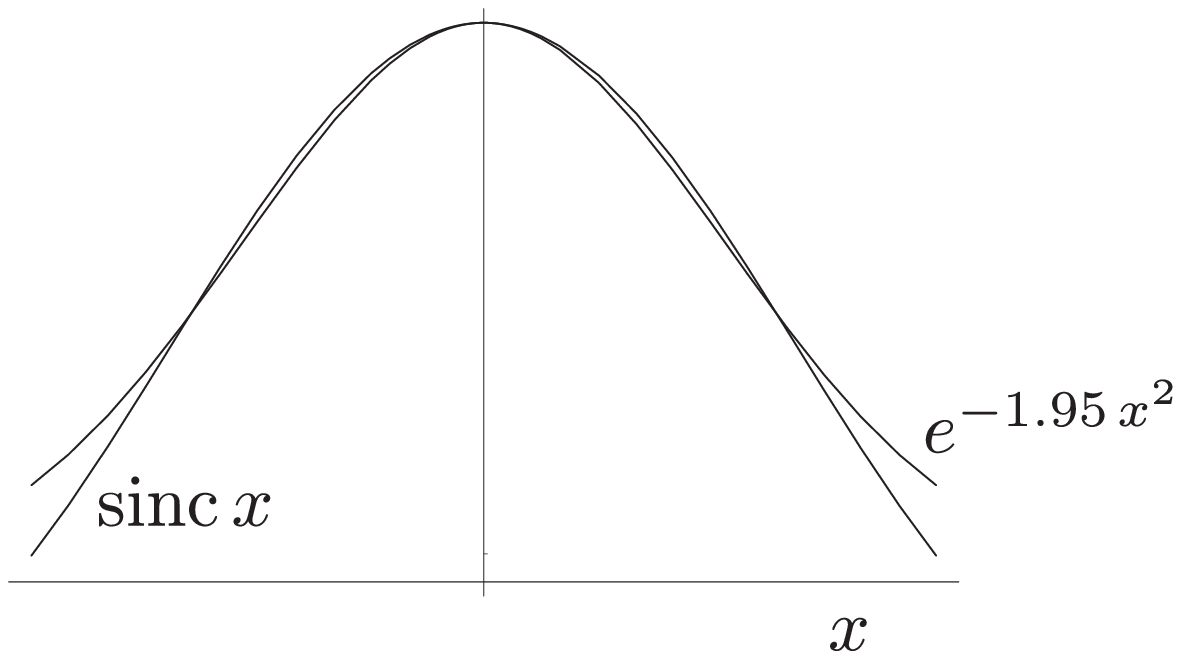}
\caption{{\protect\footnotesize {Sinc-Gauss approximation}}}
\label{Fig3}
\end{figure}

In a general case the sinc-function in Eq. (\ref{wf}) can be approximated in its main part by the Gaussian function, ${\rm sinc}\,x\approx\exp(-0.195\,x^2)$, as shown in Fig. \ref{Fig3}.

\begin{figure}[h]
\includegraphics[width=5cm]{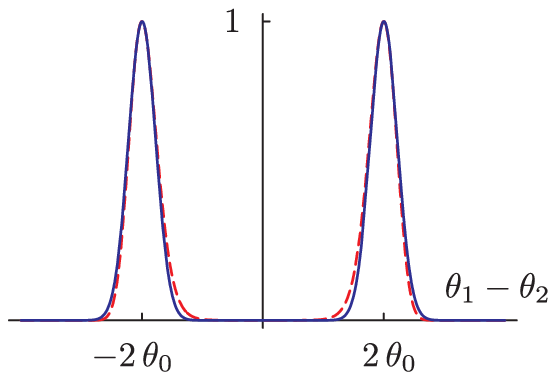}
\caption{{\protect\footnotesize {Super-Gaussian function (dashed line) and the sum of two Gaussian functions (solid line) on the left- and right-hand sides of Eq. (\ref{sum of two gausses}); $\Delta\theta_L/\theta_0\approx 0.53$}}}
\label{Fig4}
\end{figure}

\begin{figure}[h]
\includegraphics[width=6cm]{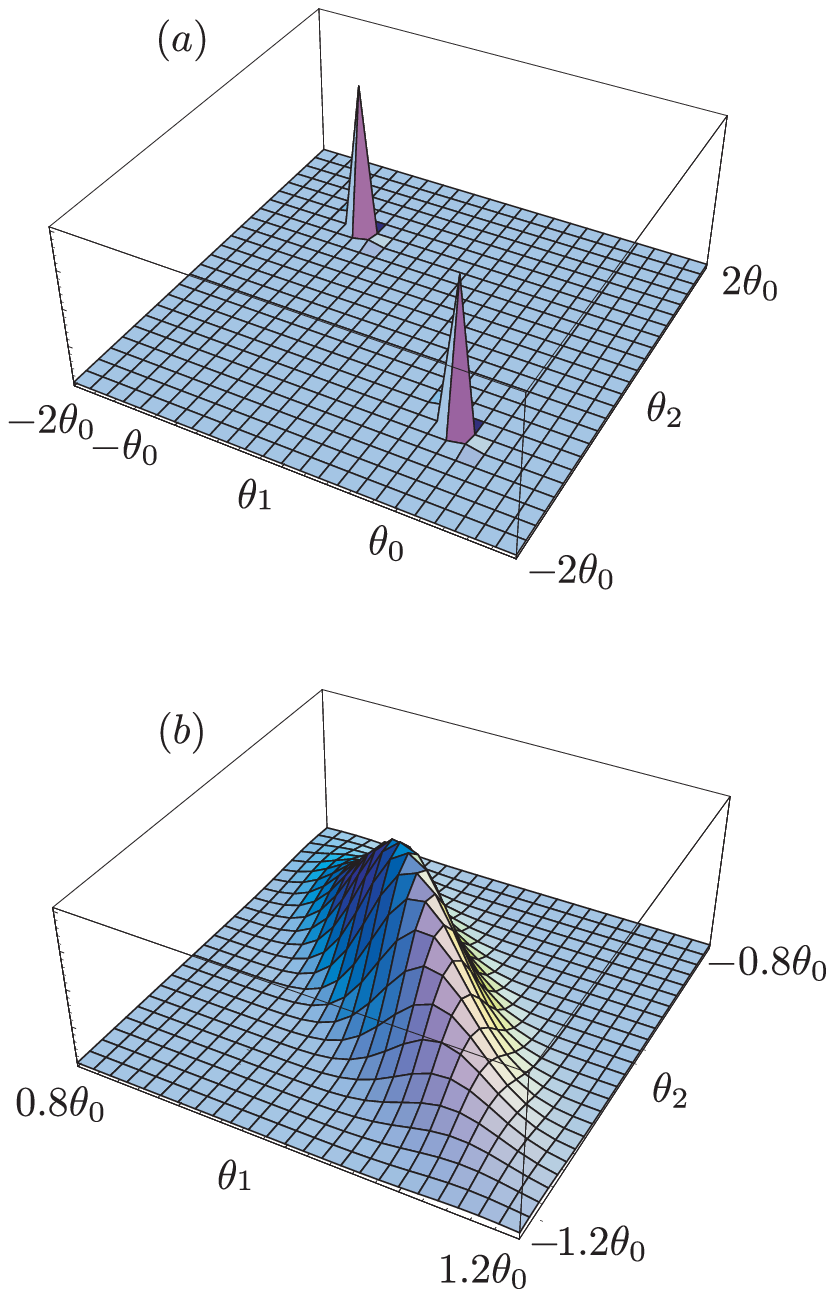}
\caption{{\protect\footnotesize {(a) Structure of the wave function  (\ref{2-double-G}) as a whole and (b) its structure in the quadrant $\{\theta_1>0,\theta_2<0\}$}}}
\label{Fig5}
\end{figure}

\noindent By applying this substitution to the wave function of Eq. (\ref{wf}) we get
\begin{equation}
 \Psi(\theta_1,\theta_2)=N\exp\left(-\frac{(\theta_1+\theta_2)^2}{2\Delta\theta_p^2}\right)
 \label{wf-G-G}
 \exp\left\{-\frac{0.195}{4\Delta\theta_L^4}\left[(\theta_1-\theta_2)^2-4\theta_0^2\right]^2\right\} .
\end{equation}
In dependence on $\theta_1-\theta_2$ the second exponent in this equation is a super-Gaussian function. As the widths  $\Delta\theta_L$ is assumed to be small compared to $\theta_0$ (\ref{narrow peaks}), in dependence on $\theta_1-\theta_2$ the super-Gaussian function in Eq. (\ref{wf}) has two well separated peaks  at $\theta_1-\theta_2=2\theta_0$ and $\theta_1-\theta_2=-2\theta_0$. For this reason the super-Gaussian function in Eq. (\ref{wf-G-G}) can be approximated very well by the sum of two true Gaussian functions
\begin{eqnarray}
 \nonumber
 \exp\left\{-\frac{0.195}{4\Delta\theta_L^4}\left[(\theta_1-\theta_2)^2-4\theta_0^2\right]^2\right\}\approx\\
 \exp\left[-\frac{0.78\,\theta_0^2}{\Delta\theta_L^4}\left(\theta_1-\theta_2-2\theta_0\right)^2\right]+
 \label{sum of two gausses}
 \exp\left[-\frac{0.78\,\theta_0^2}{\Delta\theta_L^4}\left(\theta_1-\theta_2+2\theta_0\right)^2\right].
\end{eqnarray}
A good quality of this approximation is illustrated by the picture of Fig. \ref{Fig4}.
With the substitution (\ref{sum of two gausses}) the biphoton wave function of Eqs. (\ref{wf}), (\ref{wf-G-G}) takes the form of the sum of two products of Gaussian functions
\begin{eqnarray}
 \nonumber
 \Psi(\theta_1,\theta_2)=N\exp\left(-\frac{(\theta_1+\theta_2)^2}{2\Delta\theta_p^2}\right)\times\\
 \left\{
 \exp\left[-\frac{0.78\,\theta_0^2}{\Delta\theta_L^4}\left(\theta_1-\theta_2-2\theta_0\right)^2\right]+
 \label{2-double-G}
 \exp\left[-\frac{0.78\,\theta_0^2}{\Delta\theta_L^4}\left(\theta_1-\theta_2+2\theta_0\right)^2\right]\right\}.
\end{eqnarray}
3D plots of this function are shown in two pictures of Fig. \ref{Fig5}.
The first of these pictures, \ref{Fig5}$(a)$, shows that nonzero parts of the total biphoton wave function are localized only in two quadrants in the plane $(\theta_1,\theta_2)$:  $\{\theta_1<0,\theta_2>0\}$ and $\{\theta_1>0,\theta_2<0\}$, and the photon distributions in these quadrants are perfectly symmetric. But individual distribution of each photon between the two quadrants is uncertain. The first term on the right-hand side of Eq. (\ref{2-double-G}) describes photon photon localization in the region $\theta_1\approx\theta_0$ and $\theta_2\approx -\theta_0$, whereas the second term - in the region $\theta_1\approx -\theta_0$ and $\theta_2\approx \theta_0$.

The  picture \ref{Fig5}$(b)$ shows in a more detailed scale the photon distribution inside of one of these quadrants ($\theta_1>0,\theta_2<0$). As seen clearly, this structure can be strongly asymmetric with respect to variables $(\theta_1+\theta_2)$ and $(\theta_1-\theta_2)$ in the case of significantly differing widths of biphoton distributions in the directions along $45^\circ$ and $-45^\circ$ in the ($\theta_1,\theta_2$)-plane.

As shown below, entanglement of the state (\ref{wf}), (\ref{wf-G-G}), (\ref{2-double-G}) can be determined by two reasons: uncertainty of photon localization in two quadrants of Fig. \ref{Fig5}$(a)$ and asymmetry of the angular distribution in each quadrant illustrated by the picture of Fig. \ref{Fig5}$(b)$. Analysis of section 3 took into account only the first reason, uncertainty of photon localization. For finding the degree of entanglement of the state (\ref{wf}), (\ref{wf-G-G}), (\ref{2-double-G}) as a whole and for finding its Schmidt modes we can use the same procedure as in section 3 and the same manipulations as discussed above and shown in Fig. \ref{Fig2}. As previously, the first step is changing polarizations of photons, propagating in the region around $-\theta_0$. In Eqs. (\ref{wf}), (\ref{wf-G-G}), (\ref{2-double-G}) polarization parts of the wave function are not shown. As we consider here the case of type-I phase matching, polarizations of both photons are horizontal, and in terms of polarization variables $\sigma_1$ and $\sigma_2$ the polarization wave function of two photons can be written as $\delta_{\sigma_1,H}\delta_{\sigma_2,H}$, with the numbers of polarization variables 1 and 2 associated with numbers of angular variables $\theta_1$ and $\theta_2$. This polarization wave function could be added as a factor to the angular wave function of Eqs. (\ref{wf}), (\ref{wf-G-G}), (\ref{2-double-G}). With the horizontal polarization changed for the vertical one for parts of photons moving in directions around $-\theta_0$, the total angular-polarization wave function of Eq. (\ref{2-double-G}) takes the form
\begin{eqnarray}
 \nonumber
 \Psi(\theta_1,\theta_2;\sigma_1,\sigma_2)=N\exp\left(-\frac{(\theta_1+\theta_2)^2}{2\Delta\theta_p^2}\right)\times\\
 \nonumber
 \Bigg\{\exp\left[-\frac{0.78\left(\theta_1-\theta_2-2\theta_0\right)^2}{\Delta\theta_L^4/\theta_0^2}\right]
 \delta_{\sigma_1,H}\delta_{\sigma_2,V}+\\
 \label{wf-pol-angle}
 \exp\left[-\frac{0.78\left(\theta_1+\theta_2-2\theta_0\right)^2}{\Delta\theta_L^4/\theta_0^2}\right]
 \delta_{\sigma_1,V}\delta_{\sigma_2,H}\Bigg\}.
\end{eqnarray}
Merging two beams around $\theta_0$ and $-\theta_0$ into a single beam in the first PBS of Fig. \ref{Fig2} is equivalent to the substitutions of angular variables $\theta_{1,2}\mp\theta_0\rightarrow\theta_{1,2}$ for the first term in braces of Eq. (\ref{wf-pol-angle}) and $\theta_{1,2}\pm\theta_0\rightarrow\theta_{1,2}$ for the second term, and the result is given by
\begin{eqnarray}
 \nonumber
 \displaystyle
 \Psi(\theta_1,\theta_2;\sigma_1,\sigma_2)=\sqrt{\frac{2}{\pi ab}}\,
 \displaystyle \exp\left[-\frac{(\theta_1+\theta_2)^2}{2a^2}\right]\exp\left[-\frac{(\theta_1-\theta_2)^2}{2b^2}\right]
 \times\\
 \label{wf-pol-angle-merged}
 \displaystyle\times\frac{\delta_{\sigma_1,H}\delta_{\sigma_2,V}+\delta_{\sigma_1,V}\delta_{\sigma_2,H}}{\sqrt{2}},
\end{eqnarray}
where
\begin{equation}
 \label{a-b}
 a =\frac{\theta_0}{\Delta\theta_p}, \quad b=0.8\frac{\theta_0^2}{\Delta\theta_L^2}.
\end{equation}
The angular part of the wave function (\ref{wf-pol-angle-merged}) characterizes the spreading biphoton beam with collinear central directions of propagation, for which its Schmidt decomposition is known  \cite{URen,JPB}. Specifically, with the help of Eqs. (11)-(13) of Ref. \cite{JPB} one can write down the following expansion of the angular part of the wave function (\ref{wf-pol-angle-merged}) in a series of products of the Hermite-Gaussian functions:
\begin{equation}
 \sqrt{\frac{2}{\pi ab}}\exp\left[-\frac{(\theta_1+\theta_2)^2}{2a^2}\right]\exp\left[-\frac{(\theta_1-\theta_2)^2}{2b^2}\right]
  \label{decomp-single-double-G}
 =\sum_{n=0}^\infty\sqrt{\lambda_n} \,\psi_n(\theta_1)\psi_n(\theta_2),
\end{equation}
where
\begin{equation}
 \label{Schmidt modes}
 \psi_n(x)=\frac{(2/ab)^{1/4}}{(2^nn!\sqrt{\pi})^{1/2}} \exp\left(-\frac{x^2}{ab}\right)H_n\left(\frac{\sqrt{2}\,x}{\sqrt{ab}}\right)
\end{equation}
and
\begin{equation}
 \label{lambda}
 \lambda_n=4ab\frac{(a-b)^{2n}}{(a+b)^{2(n+1)}}
\end{equation}
As mentioned above, in the case of collinear spreading beams their Schmidt modes $\psi_n$ can be separated and the parameters  of the Schmidt decomposition $\lambda_n$ can be measured experimentally \cite{Ku-Stra-1,Ku-Stra-2}. The present derivation shows that the same can be done for non-collinear spreading beams after their merging into a single beam. But this is not yet the end of story. The next step is the same as prescribed in Fig. \ref{Fig2}: the merged beam has to be sent to PBS turned for $45^\circ$ around the propagation axis to transform the wave function (\ref{wf-pol-angle-merged}) to the form
\begin{eqnarray}
 \nonumber
 \Psi= \sum_{n=0}^\infty\sqrt{\frac{\lambda_n}{2}} \Big\{\psi_n(\theta_1)\psi_n(\theta_2)\delta_{\sigma_1,45^\circ}
 \delta_{\sigma_2,45^\circ}+
 \\
 \label{Schm-angle-spreading}
 \psi_n(\theta_1+90^\circ)\psi_n(\theta_2+90^\circ)\delta_{\sigma_1,135^\circ}
 \delta_{\sigma_2,135^\circ}\Big\}.
\end{eqnarray}
After transformation we get two beams consisting of many modes each but with pairs of photons with coinciding polarizations ($45^\circ$ or $135^\circ$) propagating unsplit and in different directions. Directions of  propagation can be changed by mirrors to the original ones (around $\theta_0$ and $-\theta_0$), and polarizations of photons in each beam can be changed to the horizontal ones, after which the polarization part of the wave function ($\ket{2_H}$ for both beams) can be dropped. As the result, the Schmidt decomposition of the non-collinear angular wave function takes its final form
\begin{equation}
 \Psi(\theta_1,\theta_2)= \sum_{n=0}^\infty\sqrt{\frac{\lambda_n}{2}} \Big\{\psi_n(\theta_1-\theta_0)\psi_n(\theta_2-\theta_0)+
 \label{Schm-angle-spreading}
 \psi_n(\theta_1+\theta_0)\psi_n(\theta_2+\theta_0)\Big\}.
\end{equation}
This result shows that all terms of the Schmidt decompositions are twice degenerate, i.e., there are two pairs of Schmidt modes corresponding to each $\lambda_n$. Eigenvalues of the reduced density matrices are equal $\lambda_n/2$, and the normalization condition has the form $2\times\sum_n\lambda_n/2=1$.

Parameters characterizing the degree of entanglement of the states under consideration are the Schmidt number $K$ and the entropy of the reduced density matrix $S_r$, for which we get
\begin{equation}
 \label{K}
 K=\frac{1}{2\times\sum_n(\lambda_n/2)^2}=\frac{2}{\sum_n\lambda_n^2}=\frac{a^2+b^2}{ab}\geq 2
\end{equation}
and
\begin{equation}
 \label{Sr}
 S_r=-2\times\sum_n(\lambda_n/2)\log_2(\lambda_n/2)=1-\sum_n\lambda_n\log_2\lambda_n\geq 1.
\end{equation}
As mentioned above, angular entanglement of non-collinear angular states of biphotons arises owing to two reasons: because of asymmetry of formations in each of two occupied quadrants in the plane $(\theta_1, \theta_2)$ and because of the quadrant-quadrant symmetry of the photon distributions. The first of these two reasons occurs if $a\neq b$, and it disappears in the case $a=b$ (photon distributions in each occupied quadrant become symmetric). In this last case $\lambda_n=\delta_{n,0}$ but, because of degeneracy, the state (\ref{Schm-angle-spreading}) remains entangled with the entanglement quantifiers (\ref{K}), (\ref{Sr}) equal to $K=2$ and $S_r=1$. This remaining entanglement is related to the symmetry of two-boson wave functions, and the case $\lambda_n=\delta_{n,0}$ corresponds exactly to that of section 3.

\section{Conclusion}

Thus, the main obtained results concern derivation of the Schmidt decomposition (\ref{Schm-angle-spreading}) for the angular wave function of the non-collinear frequency-degenerate biphoton states (\ref{wf}) arising in the SPDC process with the type-I phase matching and angular widths of photon distribution completely taken into account. The derivation is based on the proposal of an experiment which can provide separation of Schmidt modes and measurement of parameters of the Schmidt decomposition. The  derivation and the proposed experiment consist of three steps: (1) manipulation with photon polarization which provides duplication of the symmetry of angular wave functions by the symmetry of polarization states, (2) transformation/merging of the pair of non-collinear photon beams to a single collinear one with the same degree of entanglement and the same amount of modes, and (3) the polarization-sensitive splitting/unmerging of the collinear beam into a pair of beams with photons regrouped in such a way that each of two unmerged beams contains only unsplit pairs of photons. Uncertainty of localization of photon pairs in these new beams is responsible for the entanglement related to symmetry of biphoton states, and a true  angular entanglement is determined by the amounts of non-zero terms in the decomposition of angular wave functions in each of two channels into sums of products of Hermite-Gaussian one-photon angular functions (Schmidt modes). In experiment separation of terms corresponding to different products of Hermite-Gaussian angular functions can be performed at the stage of a merged beam between two PBS  in the scheme of Fig. \ref{Fig2}.

Note that another type of experiment can be related to measuring coincidence and single-particle angular distributions of photons. In a scheme with two detectors one can install one of them for counting photons moving strictly in the direction $-\theta_0$ with the second detector scanning around the direction $\theta_0$. The coincidence distribution found in this way is characterized by its width (e.g., the FWHM width) $\Delta\theta_1^{(c)}$. Measurements with the turned off detector at $-\theta_0$ will give a single-particle distribution $\Delta\theta_1^{(s)}$. In accordance with the idea of Ref. \cite{2004} the ratio of these widths $R^{(\rm part1)}=\Delta\theta_1^{(s)}/\Delta\theta_1^{(c)}$ can be considered as a measure of the degree of entanglement. For non collinear beams this will be a partial entanglement, e.g., for the part of the biphoton angular wave function located in the quadrant $(\theta_1>0,\theta_2<0)$ in Fig. \ref{Fig5}$(a)$. As known \cite{2006}, for double-Gaussian bipartite wave functions the parameter $R^{(\rm part1)}$ coincides exactly with the corresponding Schmidt number $K^{(\rm part1)}$. If one changes the roles of the detectors by keeping constant the position of the detector at $\theta_0$ and scanning the second detector around the direction $-\theta_0$, one can measure the widths $\Delta\theta_2^{(c)}$ and $\Delta\theta_2^{(s)}$, as well as the parameters $R^{(\rm part2)}=K^{(\rm part2)}$ for the second occupied quadrant of the angular wave function, $(\theta_1>0,\theta_2<0)$ in Fig. \ref{Fig5}$(a)$. Because of the symmetry of biphoton wave functions, the partial Schmidt numbers $K^{(\rm part1)}$ and $K^{(\rm part2)}$ must be equal to each other. Thus, in terms of the Schmidt number $K$, the total degree of entanglement of the angular wave function as a whole is determined by the sum of partial contributions, $K=K^{(\rm part1)}+K^{(\rm part2)}=2K^{(\rm part1)}\geq 2$.

It would be very interesting to perform two types of experiments together: ($a$) finding  parameters $\lambda_n$ of the Schmidt decomposition (\ref{Schm-angle-spreading}) as described above and determining the Schmidt number $K$ via Eq. (\ref{K}) and ($b$) finding the same entanglement quantifier $K$ by means of measurements of the widths of the coincidence and single-particle distributions to find the width-ratio parameters $R^{(\rm part1,2)}$ and identifying their sum with the Schmidt number $K$. Comparison of results of these two experiments can be very interesting.

\section*{Acknowledgement}
The work is supported by the Russian Science Foundation, grant  14-12-01338

\end{document}